\documentclass[aps,prl, twocolumn, superscriptaddress]{revtex4}
\usepackage{graphicx}
\usepackage{bm}

\begin{document}

\title
{ Exotic vs. conventional scaling and universality in a
\\ disordered bilayer quantum Heisenberg antiferromagnet }

\author{Rastko Sknepnek}
\author{Thomas Vojta}
\affiliation{Physics Department, University of Missouri - Rolla, Rolla, MO 65409}
\author{Matthias Vojta}
\affiliation{\mbox{Institut f{\" u}r Theorie der Kondensierten Materie, Universit{\" a}t
Karlsruhe, 76128 Karlsruhe, Germany}}
\date{\today}

\begin{abstract}
We present Monte-Carlo simulations of a two-dimensional (2d) bilayer quantum Heisenberg
antiferromagnet with random dimer dilution. In contrast to exotic scaling scenarios found in
other random quantum systems, the quantum phase transition in this system is characterized by a
finite-disorder fixed point with power-law scaling. After accounting for corrections to
scaling, with a leading irrelevant exponent of $\omega\approx0.48$, we find universal critical
exponents $z=1.310(6)$ and $\nu=1.16(3)$. We discuss the consequences of these findings and
suggest new experiments.
\end{abstract}

\maketitle


Quantum phase transitions (QPT) under the influence of quenched disorder are a topic of great
current interest. Experimental examples range from localized \cite{magnets} and itinerant
\cite{ditusa} quantum magnets to heavy-fermion compounds \cite{hf}, high-temperature
superconductors \cite{cuprate}, and to metal-insulator \cite{mit} and superconductor-insulator
transitions \cite{scit}. These systems display rich new physics but many are still poorly
understood. In the context of classical phase transitions, the interplay between disorder and
critical fluctuations has a long history. Harris \cite{harris} derived a criterion for the
stability of a critical point against disorder: If the correlation length exponent $\nu$
fulfills the inequality $\nu>2/d$, where $d$ is the spatial dimensionality, the critical
behavior is not influenced by weak disorder. If a clean critical point violates the Harris
criterion, the generic result of introducing disorder is a new (finite-disorder) critical point
with power-law scaling and new critical exponents which fulfill the Harris criterion
\cite{chayes}.

At QPTs, order-parameter fluctuations in space and time must be considered. Quenched
disorder is perfectly correlated in time direction. As a result, disorder effects at QPTs
are generically stronger than at classical transitions. Prominent consequences are the
infinite-randomness critical points in 1d random spin chains \cite{chains} and in 1d
\cite{ising1d} and 2d \cite{ising2d,motrunich} random quantum Ising models. At these
critical points, the dynamical scaling is activated, i.e., correlation time $\xi_\tau$
and correlation length $\xi$ obey $\ln \xi_\tau\sim \xi^\mu$. (At conventional critical
points, this relation is a power law, $\xi_\tau\sim \xi^z$, with a universal dynamical
exponent $z$).
In itinerant electron systems, the effects of impurities can be even more dramatic. For
Ising symmetry, the interplay of quenched disorder and Landau damping of the order
parameter fluctuations completely destroys the sharp QPT by smearing \cite{vojta}.
Further exotic phenomena include non-universal, continuously varying exponents, observed
in the Griffiths region associated with a QPT \cite{Griffiths,ising1d,ising2d} or at
certain impurity QPTs \cite{kondo}. On the other hand, the {\em stable low-energy} (as
opposed to critical) fixed point of random Heisenberg models in $d\ge2$ has been shown to
be conventional \cite{igloi2003}. Preliminary results \cite{motrunich} for the {\em
critical} point in these models suggest that the infinite-randomness fixed point is
unstable, but no definite conclusion on the fate of the transition has been reached.
These results lead to the general question whether all QPTs in presence of quenched
disorder are unconventional.

In this Letter, we provide a ``proof of principle'' that this is not the case: The QPT of
a dimer-diluted spin-1/2 bilayer quantum Heisenberg antiferromagnet is shown to exhibit a
conventional finite-disorder critical point with power-law dynamical scaling and
universal critical exponents. After accounting for corrections to scaling characterized
by an irrelevant exponent $\omega \approx 0.48$ we find the asymptotic dynamical and
correlation length exponents to be $z=1.310(6)$ and $\nu=1.16(3)$ (fulfilling the Harris
criterion $\nu>2/d=1$ \cite{harris,chayes}).

\begin{figure}
\centerline{
\includegraphics[width=0.85\columnwidth]{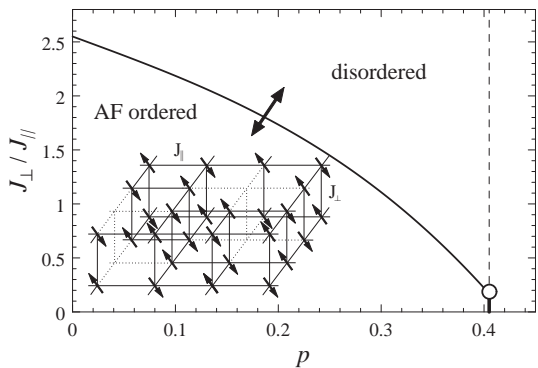}
}
\vspace*{-10pt}
\caption{ Phase diagram \cite{vajk} of the diluted bilayer Heisenberg antiferromagnet, as
function of $J_\perp/J_\parallel$ and dilution $p$. The dashed line is the percolation
threshold, the open dot is the multicritical point of Refs.~\protect\cite{sandvik,vajk}. The
arrow indicates the QPT studied here. Inset: The model: Quantum spins (arrows) reside on the
two parallel square lattices. The spins in each plane interact with the coupling strength
$J_\parallel$. Interplane coupling is $J_\perp$. Dilution is done by removing dimers.}
\label{fig:model}
\end{figure}

Our starting point is a bilayer quantum Heisenberg antiferromagnet as depicted in the inset of
Fig.\ \ref{fig:model}. The spins in each 2d layer interact via nearest neighbor exchange
$J_\parallel$, and the interplane coupling is $J_\perp$.  The clean version of this model has
been studied extensively \cite{clean_bilayer1, clean_bilayer2}. For $J_\perp\gg J_\parallel$,
neighboring spins from the two layers form singlets, and the ground state is paramagnetic. In
contrast, for $J_\parallel\gg J_\perp$ the system develops N\'eel order. Both phases are
separated by a QPT at $J_\perp / J_\parallel \approx 2.525$. Random disorder is introduced by
removing {\em pairs} (dimers) of adjacent spins, one from each layer. The Hamiltonian of the
model with dimer dilution is:
\begin{equation}
\label{eq:1} H=J_\parallel \sum_{{\langle i,j\rangle} \atop a=1,2}\epsilon_i\epsilon_j
{\mathbf{\hat{S}}}_{i,a}\cdot{\mathbf{\hat{S}}}_{j,a}+J_\perp\sum_i\epsilon_i{\mathbf{\hat{S}}}_{i,1}\cdot{\mathbf{\hat{S}}}_{i,2},
\end{equation}
and $\epsilon_i$=0 ($\epsilon_i$=1) with probability $p$ ($1-p$).

The phase diagram of the dimer-diluted bilayer Heisenberg model has been studied by Sandvik
\cite{sandvik} and Vajk and Greven \cite{vajk}, see Fig.~\ref{fig:model}. For small $J_\perp$,
magnetic order survives up to the percolation threshold $p_p\approx 0.4072$, and a
multicritical point exists at $p=p_p$ and $J_\perp / J_\parallel \approx 0.16$. We focus on the
generic transition at $0<p<p_p$, driven by $J_\perp$, where the results of
Refs.~\cite{sandvik,vajk} are inconclusive.

To determine the critical behavior at the QPT effectively, we proceed by mapping the
quantum Hamiltonian (\ref{eq:1}) onto a classical model. First we note that the
low-energy properties of bilayer quantum antiferromagnets are represented by a
(2+1)-dimensional O(3) quantum rotor model \cite{nlsm} with the rotor coordinate
$\mathbf{\hat{n}}_i$ corresponding to $\mathbf{\hat{S}}_{i,1} - \mathbf{\hat{S}}_{i,2}$
and the angular momentum $\mathbf{\hat{L}}_i$ representing $\mathbf{\hat{S}}_{i,1} +
\mathbf{\hat{S}}_{i,2}$ (see, e.g., chapter 5 of Ref. \cite{book}). This quantum rotor
model in turn is equivalent to a 3d classical Heisenberg model with the disorder
perfectly correlated in imaginary time direction, as can be easily seen from a path
integral representation of the partition function. Thus, our classical Hamiltonian reads:
\begin{equation}
\label{eq:2} H=K\sum_{\langle i,j\rangle, \tau}
\epsilon_i\epsilon_j\mathbf{n}_{i,\tau}\cdot\mathbf{n}_{j,\tau}+K\sum_{i,\tau}\epsilon_i\mathbf{n}_{i,\tau}\cdot\mathbf{n}_{i,\tau+1},
\end{equation}
where $\mathbf{n}_{i,\tau}$ is an O(3) unit vector. The coupling constant $\beta K$ of
the classical model is related to the ratio $J_\parallel / J_\perp$ of the quantum model.
Here, $\beta\equiv1/T$ where $T$ is an effective ``classical'' temperature, not equal to
the real temperature which is zero. We set $K=1$ and drive the classical system through
the transition by tuning the classical temperature $T$.

As an aside, we note that dimer dilution in the quantum model (\ref{eq:1}) does not
introduce random Berry phases because the Berry phase contributions from the two spins of
each unit cell cancel \cite{nlsm,book}. In contrast, for site dilution, the physics
changes completely: The random Berry phases (which have no classical analogue) are
equivalent to impurity-induced moments \cite{ssmv}, and those become weakly coupled via
bulk excitations. Thus, for all $p<p_p$ the ground state shows long-range order,
independent of $J_\perp / J_\parallel$! This effect is absent for dimer dilution, and
both phases of the clean system survive for small $p$ \cite{berry}.

The classical model (\ref{eq:2}) is studied by Monte-Carlo simulations using the
efficient Wolff cluster algorithm \cite{wolff, frustration}. We investigate linear sizes
up to $L=120$ in space direction and $L_\tau=384$ in imaginary time, for impurity
concentrations $p=\frac{1}{8}$, $\frac 1 5$, $\frac 2 7$ and $\frac 1 3$. The results are
averaged over $10^3$ -- $10^4$ disorder realizations. Each sample is equilibrated using
100 Monte-Carlo sweeps (spin-flips per site). For large dilutions, $p=\frac 2 7$ and
$\frac 1 3$ we perform both Wolff and Metropolis sweeps to equilibrate small dangling
clusters. During the measurement period of another 100-200 sweeps we calculate
magnetization, susceptibility, specific heat and correlation functions.

A quantity particularly suitable to locate the critical point and to extract high precision
values for the exponents $z$ and $\nu$ is the Binder ratio:
\begin{equation}
\label{eq:3} g_{av}=\left[ 1-\frac{\langle |\mathbf{M}|^4\rangle}{3\langle
|\mathbf{M}|^2\rangle^2}\right]_{av},
\end{equation}
where $\mathbf{M}=\sum_{i,\tau}\mathbf{n}_{i.\tau}$, $\left[\ldots\right]_{av}$ denotes the
disorder average and $\langle\ldots\rangle$ denotes the Monte-Carlo average for each sample.
This quantity has scale dimension 0. Thus, its finite-size scaling form is given by
\begin{eqnarray}
\label{eq:4} g_{av}&=&\tilde{g}_C (tL^{1/\nu},L_\tau/L^z)\qquad {\rm or}
\\ \label{eq:5} g_{av}&=&\tilde{g}_A (tL^{1/\nu},\log(L_\tau)/L^\mu)
\end{eqnarray}
for conventional scaling or for activated scaling, respectively. Two important
characteristics follow: (i) For fixed $L$, $g_{av}$ has a peak as a function of
$L_{\tau}$. The peak position $L_{\tau}^{\rm max}$ marks the {\em optimal} sample shape,
where the ratio $L_{\tau}/L$ roughly behaves like the corresponding ratio of the
correlation lengths in time and space directions, $\xi_{\tau}/\xi$. At the critical
temperature $T_c$, the peak value $g_{av}^{\rm max}$ is independent of $L$. Thus, for
power law scaling, plotting $g_{av}$ vs. $L_\tau/L_\tau^{max}$ at $T_c$ should collapse
the data, without the need for a value of $z$. In contrast, for activated scaling the
$g_{av}$ data should collapse when plotted as a function of
$\log(L_\tau)/\log(L_\tau^{\rm max})$. (ii) For samples of the optimal shape
($L_\tau=L_\tau^{max}$), plots of $g_{av}$ vs. temperature for different $L$ cross at
$T_c$. Based on these two characteristics, we use a simple iterative procedure to
determine both the optimal shapes and the location of the critical point.

\begin{figure}
\includegraphics[width=\columnwidth]{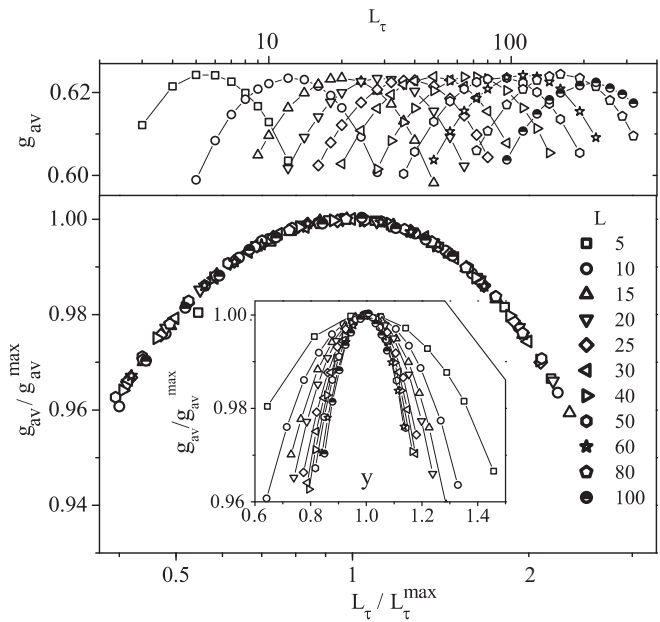}
\caption{ Upper panel: Binder ratio $g_{av}$ as a function of $L_\tau$ for various $L$
($p=\frac 1 5$). Lower panel: Power-law scaling plot $g_{av}/g_{av}^{max}$ vs.
$L_\tau/L_\tau^{max}$  Inset: Activated scaling plot $g_{av}/g_{av}^{max}$ vs.
$y=\log(L_\tau)/\log(L_\tau^{max})$.  } \label{fig:zscaling}
\end{figure}

We now turn to our results. To distinguish between activated and power-law dynamical
scaling we perform a series of calculations at the critical temperature. The upper panel
of Fig.\ \ref{fig:zscaling} shows the Binder ratio $g_{av}$ as a function of $L_\tau$ for
various $L=5\ldots 100$ and dilution $p=\frac 1 5$ at $T=T_c=1.1955$. The statistical
error of $g_{av}$ is below $0.1\%$ for the smaller sizes and not more than $0.2\%$ for
the largest systems. As expected at $T_c$, the maximum Binder ratio for each of the
curves does not depend on $L$.
To test the conventional power-law scaling form, eq.\ (\ref{eq:4}), we plot
$g_{av}/g_{av}^{max}$ as a function of $L_\tau/L_\tau^{max}$ in the lower panel of Fig.\
\ref{fig:zscaling}. The data scale extremely well, giving statistical errors of
$L_\tau^{\rm max}$ in the range between $0.3\%$ and $1\%$.  For comparison, the inset
shows a plot of $g_{av}$ as a function of $\log(L_\tau)/\log(L_\tau^{max})$ corresponding
to eq.\ (\ref{eq:5}). The data clearly do not scale which rules out the activated scaling
scenario.
The results for the other impurity concentrations $p=\frac 1 8, \frac 2 7, \frac 1 3$ are
completely analogous.

\begin{figure}
\includegraphics[width=0.8\columnwidth]{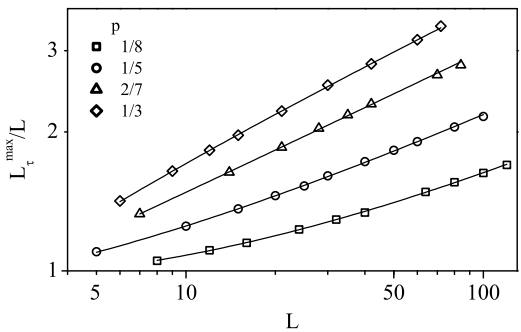}
\caption{ $L_\tau^{max}/L$ vs. $L$ for four disorder concentrations $p = \frac 1 8$, $\frac 1
5$, $\frac 2 7$ and $\frac 1 3$. Solid lines: Fit to $L_\tau^{max}=aL^z(1+bL^{-\omega_1})$ with
$z=1.310(6)$ and $\omega_1=0.48(3)$.  } \label{fig:zfit}
\end{figure}

Having established conventional power-law  dynamical scaling, we proceed to determine the
dynamical exponent $z$. In Fig.\ \ref{fig:zfit}, we plot $L_\tau^{max}$ vs. $L$ for all
four dilutions $p$.
The curves show significant deviations from pure power-law behavior which can be
attributed to corrections to scaling due to irrelevant operators. In such a situation, a
direct power-law fit of the data will only yield {\em effective} exponents. To find the
true {\em asymptotic} exponents we take the leading correction to scaling into account by
using the ansatz $L_\tau^{max}(L)=aL^z(1+bL^{-\omega_1})$  with universal
(dilution-independent) exponents $z$ and $\omega_1$ but dilution-dependent $a$ and $b$. A
combined fit of all four curves gives $z=1.310(6)$ and $\omega_1=0.48(3)$ where the
number in brackets is the standard deviation of the last given digit. The fit is of high
quality ($\chi^2\approx 0.7$) and robust against removing complete data sets or removing
points form the lower or upper end of each set. We thus conclude that the asymptotic
dynamical exponent $z$ is indeed universal. (Note that the leading corrections to scaling
vanish very close to $p=\frac 2 7$; the curvature of the $L_\tau^{\rm max}(L)$ curves in
Fig.\ \ref{fig:zfit} is opposite above and below this concentration.)

\begin{figure}
\includegraphics[width=0.8\columnwidth]{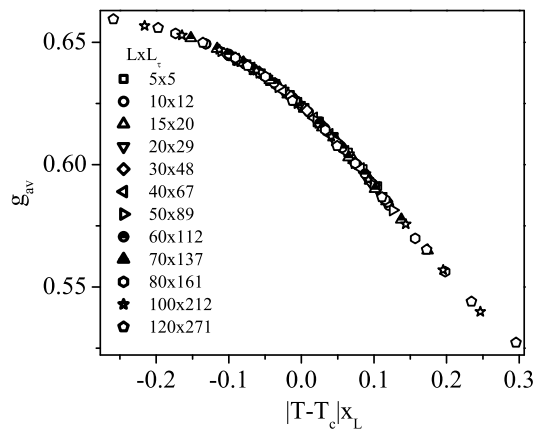}
\caption{Scaling plot of $g_{av}$ vs. $(T-T_c)x_L$ for $p=0.2$.
   $x_L$ is the factor necessary to scale the data onto a master curve.}
\label{fig:nuscaling}
\end{figure}

\begin{figure}
\includegraphics[width=0.8\columnwidth]{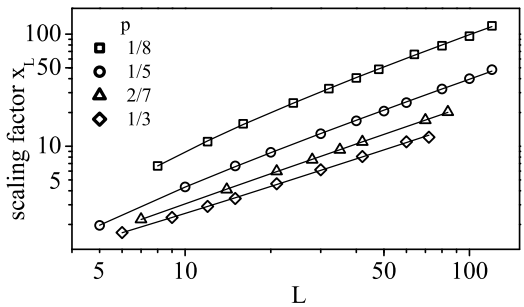}
\caption{Scaling factor vs. $L$ for four disorder concentrations $p = \frac 1 8$, $\frac 1 5$,
$\frac 2 7$ and $\frac 1 3$. Solid lines: Fit to $x_L=cL^{1/\nu}(1+dL^{-\omega_2})$ with
$\nu=1.16(3)$ and $\omega_2=0.5(1)$.  } \label{fig:nufit}
\end{figure}

To find the correlation length exponent $\nu$, we perform simulations in the vicinity of $T_c$
for samples with the optimal shape ($L_\tau=L_\tau^{max}$) to keep the second argument of the
scaling function (\ref{eq:4}) constant. Fig.\ \ref{fig:nuscaling} shows a scaling plot of
$g_{av}$ vs.\ $T$ for impurity concentration $p = \frac 1 5$.
Again, the data scale very well, but since the scaling function lacks the characteristic
maximum, the error of the resulting scaling factor $x_L$ is somewhat larger ($1\ldots2\%$) than
that of $L_\tau^{\rm max}$. The same quality of scaling was achieved for the other dilutions.
Fig.~\ref{fig:nufit} shows the scaling factor $x_L$ vs.\ $L$ for all four data sets.
A combined fit to the ansatz $x_L=cL^{1/\nu}(1+dL^{-\omega_2})$ where $\nu$ and $\omega_2$ are
universal, gives $\nu=1.16(3)$ and $\omega_2=0.5(1)$. As above, the fit is robust and of high
quality ($\chi^2 \approx 1.2$). Importantly, as expected for the true asymptotic exponent,
$\nu$ fulfills the Harris criterion \cite{harris}, $\nu>2/d$=1. Note that both irrelevant
exponents $\omega_1$ and $\omega_2$ agree within their error bars, suggesting that the same
irrelevant operator controls the leading corrections to scaling for both $z$ and $\nu$.

We have also calculated total magnetization and susceptibility. The corresponding
exponents $\beta/\nu=0.56(5)$ and $\gamma/\nu=2.15(10)$ have slightly larger error bars
than $z$ and $\nu$. Nonetheless, they fulfill the hyperscaling relation
$2\beta+\gamma=(d+z)\nu$ which is another argument for our results being asymptotic
rather than effective exponents.

In summary, we have performed Monte-Carlo simulations of a 3d classical Heisenberg model
with linear impurities which is in the same universality class as the dimer-diluted
bilayer quantum Heisenberg antiferromagnet. We have shown that the QPT in this system is
controlled by a conventional, finite-disorder critical point with power-law dynamical
scaling and universal exponents. (Note that the Ising version of our model, the diluted
2d random transverse Ising model, shows an infinite-randomness critical point
\cite{ising2d,senthil}.)

Let us compare our results to previous work. The multicritical point at $p=p_p$ and
$J_\perp / J_\parallel \approx 0.16$, found in Refs.~\cite{sandvik,vajk}, has a dynamical
exponent $z\approx1.3$. Within the error bars, this value coincides with the one found
here for the generic $p<p_p$ transition. We see no a-priori reason for this coincidence,
so far it is unclear whether or not it is accidental.  Vajk and Greven \cite{vajk} also
quote exponents for $p<p_c$. At dilution $p=0.25$ they find $z=1.07$ and $\nu=0.89$,
different from our results. However, as the authors of Ref.\ \cite{vajk} point out, a
value of $\nu<1$ violates the Harris criterion, indicating that it represents an
effective rather than an asymptotic exponent. It would also be useful to compare our
findings with analytical results. To the best of our knowledge, the only quantitative
result is a resummation of the 2-loop $\epsilon$-expansion \cite{ferber}. The predicted
exponents significantly differ from ours; but they also violate the Harris criterion,
casting doubt on their validity.

Finally, we comment on experiments. If chemical doping replaces magnetic by non-magnetic ions
in an antiferromagnet, e.g., Cu by Zn in YBa$_2$Cu$_3$O$_6$, the case of site rather than dimer
dilution is realized. The most promising way to achieve bond dilution is the introduction of
strong antiferromagnetic intra-dimer bonds at random locations. Thus we propose to study
magnetic transitions in bond-disordered systems; those transitions can be expected to be in the
same universality class as the one studied here. One candidate material -- albeit 3d -- is
(Tl,K)CuCl$_3$ \cite{tlkcucl} under pressure; interesting quasi-2d compounds are
SrCu$_2$(BO$_3$)$_2$ or BaCuSi$_2$O$_6$, where suitable dopants remain to be found.

We acknowledge partial support from the University of Missouri Research Board, from the
NSF under grant No. DMR-0339147 and from the DFG Center for Functional Nano\-struc\-tures
Karls\-ruhe.



\begin{thebibliography}{99}
\bibitem{frustration}
Since the classical model is not frustrated, we can use cluster algorithms to reduce critical
slowing down.
\end{thebibliography}
\end{document}